\documentstyle[aps,multicol,epsfig,psfig]{revtex}   

\newcommand{\beq}{\begin{equation}}  
\newcommand{\eeq}{\end{equation}}  
\newcommand{\bea}{\begin{eqnarray}}  
\newcommand{\eea}{\end{eqnarray}}  
  
\begin{document}     
\title{Multielectron ionization of atoms in strong fields:\\
Classical analysis in symmetric subspace}
\author{
Krzysztof Sacha$^{1,2}$ and Bruno Eckhardt$^1$
}     
\address{$^1$ Fachbereich Physik, Philipps Universit\"at    
	 Marburg, D-35032 Marburg, Germany}    
\address{$^2$ Instytut Fizyki im. Mariana Smoluchowskiego,
Uniwersytet Jagiello\'nski, ul. Reymonta 4,
PL-30-059 Krak\'ow, Poland}
\date{\today}     
\maketitle{ }   
  
\begin{abstract}
We consider the final stage of multiple ionization of atoms in a 
strong linearly polarized laser field within a classical model.
We propose that non-sequential multiple ionization is dominated
by symmetric escape from a highly excited intermediate complex.
For a configuration of $N$ electrons with $C_{Nv}$ symmetry in 
a plane perpendicular to the electric field we analyze the classical
motion in phase space and discuss the final momentum distribution 
parallel and perpendicular to the polarization axis. 
The results are in good agreement with recent experiment of
multiple ionization in Ne. 
\end{abstract}     
  
\begin{multicols}{2}  
Double (or multiple) ionization of atoms is a fundamental process to
our understanding of many electron dynamics in external fields.
 The surprisingly large yields of multiply charged ions 
 reported in the first experiments with intense laser fields
 \cite{lhuiller} clearly show
that sequential ionization is not the leading process
and that the
electron-electron correlated dynamics has to be taken into account. 
One possibility is that the 
electron is shaken-off due to a non-adiabatic change in potential 
during the ionization of the first electron \cite{Kulander1},
a process that accounts for double ionization at high photon energies
(above 1 keV) \cite{schmidt}.
For multiphoton ionization 
a two step process is more likely: one electron is ionized first,
accelerated by the field and driven back to the atom where it
ionizes the second electron in a rescattering collision
\cite{Corkum,Kulander3,becker1,becker2,Kulander5,becker3,becker4}.
The applicability of tunneling expressions as in \cite{Corkum,eichmann}
then is due to the first ionization taking place through 
a tunneling process in a quasi-static electric field.
Such a sequence of events is supported also by the numerical 
calculations of Becker and Faisal \cite{becker4}.

The correlation between the electrons was 
convincingly demonstrated in a series of
recent experiments. Measurements of the 
distributions of ion and electron momenta in double and
triple ionization \cite{weber1,weber2,weber3,rottke} clearly show a
preference for electrons escaping towards the
same side of the atomic core along the field polarization axis\cite{weber3}. 
On first sight this seems to be very different from the symmetric
escape of electrons in double ionization in the absence of a field,
where according to Wanniers analysis the electrons escape in
opposite directions \cite{Wannier,Rau}. A closer analysis of 
Wanniers arguments shows, however, that they can be applied to this situation
as well \cite{eckhardt00}. It is our aim here to generalize this model to 
the correlated escape of three or more electrons.

The essential elements of Wanniers analysis are a division of the 
process into two steps, the formation of a highly excited complex
of two electrons close to the core and the two electron escape
from this complex. At the threshold for ionization, energy is
scarce. Therefore, in the dominant channel,
mutual repulsion should be minimal and the energy
should be equally shared by the outgoing electrons. Any deviation
from the symmetric arrangement in phase space or in energy would
be amplified and lead to single rather than multiple ionization.
These requirements become less stringent the higher the energy 
above threshold.

One can argue that a similar division 
of the ionization process is possible in electric fields\cite{eckhardt00}.
In particular, ionization also proceeds through the formation of 
an intermediate highly excited complex, created during the rescattering 
event. But the next step is different, the configuration that 
dominates the ionization channel is no longer the Wannier state
and the threshold energy is modified as well. 
For instance, double ionization is observed even though
the estimated energy transferred by returning electron is too small for 
immediate double ionization \cite{weber1}. As discussed in \cite{eckhardt00}
this is possible if during the collision the external field is not zero,
for then a Stark saddle opens through which the electrons can escape. 
This saddle breaks the symmetry and focuses the electrons in 
the direction of the electric field.  
It also forces the electrons to be close to a symmetry subspace, since
by mutual repulsion the electron that reaches the 
saddle first can push the other back to the nucleus. This
would then result in either single ionization 
or in another rescattering event, but not in double ionization.
Classical trajectory calculations within this symmetry subspace are
able to reproduce the main features of the experimentally observed 
ion momentum distributions.

In the present letter we present a generalization of our model
\cite{eckhardt00} to multielectron ionization. The key assumption
is that the process is dominated by a symmetric configuration of the 
electrons with respect to the field polarization axis
as suggested by the experiments \cite{weber1,weber2,weber3,rottke}. 
Specifically,
we assume that all electrons move in a plane perpendicular to the
field and that they obey a $C_{Nv}$ symmetry, which generalizes the
$C_{2v}$ symmetry of the previously analyzed case.
The reflection symmetry limits the momenta to be parallel to the 
symmetry planes and thus confines the motion to a dynamically allowed 
subspace in the high-dimensional $N$-body phase space. I.e. 
if in the full phase space of the $N$-body problem initial
conditions are prepared in this subspace they will never leave it.

We take the electric field directed along the $z$-axis, and the 
positions of the $N$-electrons as 
$z_i=z$, $\rho_i=\rho$ and $\varphi_i=2\pi i/N$,
where $(\rho_i,\varphi_i,z_i)$ are cylindrical coordinates. 
The momenta of the electrons are all identical, 
$p_{\rho,i}=p_\rho$, $p_{z,i}=p_z$ and $p_{\varphi,i}=0$.
For this geometry the classical 
Hamiltonian for $N$ electrons,
for zero total angular momentum along the field axis, in
atomic units, with infinitely heavy nucleus and in dipole approximation,
reads
\beq
H(p_\rho,p_z,\rho,z,t)=N
\frac{p_\rho^2+p_z^2}{2}+V(\rho,z,t),
\label{h}
\eeq
with potential energy
\beq
V
=-\frac{N^2}{\sqrt{\rho^2+z^2}}+\frac{N(N-1)}{4\rho\sin (\pi/N)}
+NzFf(t)\cos (\omega t+\phi)
\label{p}
\eeq
and the pulse shape
\beq
f(t)=\sin^2(\pi t/T_d). 
\label{ps}
\eeq
$F$, $\omega$ and $\phi$ stand for peak amplitude, frequency and phase of the
external field respectively, while $T_d$ is the pulse duration.

Experiments show that the multiple ionization is possible
even when the energy transferred by the rescattered electron is less
than that needed to lift the system to the many electron continuum.
We therefore assume that the complex from which multiple ionization
starts is described by Hamiltonian (\ref{h}) with initially negative
energy $E$. Moreover, since the electronic motion close to the nucleus
is much faster than the changes in the electric field 
the process of electron escape can
be discussed adiabatically for fixed external field.
This allows us to analyze the potential energy for a
fixed time. 

Before we proceed with the analysis we take advantage of the
scaling symmetry of the classical Hamiltonian (\ref{h}) and eliminate
one parameter. If the variables are rescaled according to
\bea
H&\rightarrow& \sqrt{F}H, \cr
\rho,\ z&\rightarrow& \rho/\sqrt{F},\ z/\sqrt{F}, \cr
p_\rho,\ p_z&\rightarrow& F^{1/4}p_\rho,\ F^{1/4}p_z, \cr
\omega,\ t&\rightarrow& F^{3/4}\omega,\ F^{-3/4}t
\label{sc}
\eea
the dynamics becomes independent of the peak value of the 
field amplitude, i.e. we obtain the system described by the 
Hamiltonian (\ref{h}) with $F=1$.

\begin{figure}[hbt]
\epsfig{file=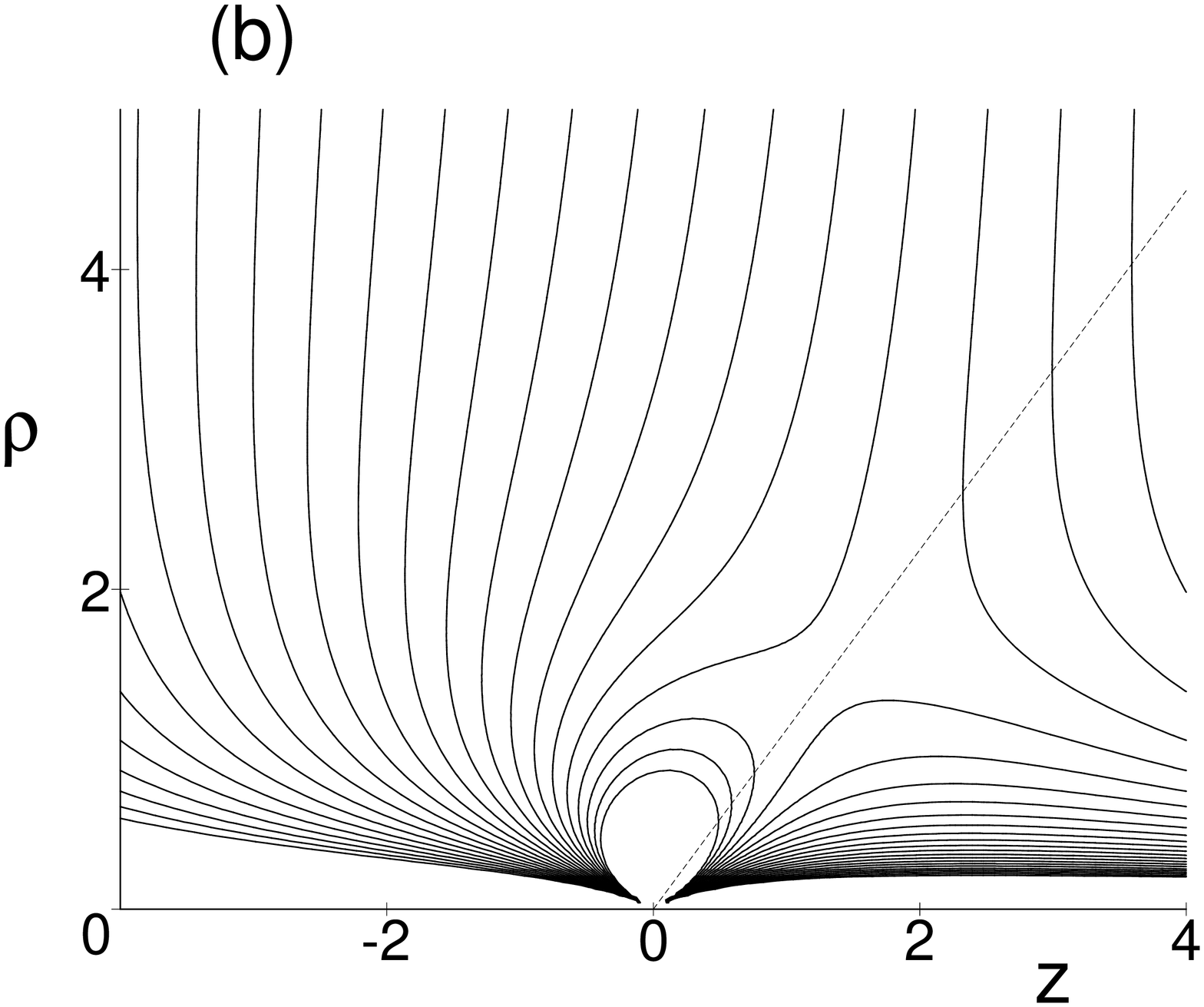%
,scale=0.34,angle=-90
}
\caption[]{
Equipotential curves of the potential energy (\protect{\ref{p}}),
in the scaled variables (\protect{\ref{sc}}), for two different values of the
number of electrons: $N=3$ [panel (a)] and $N=6$ [panel (b)]. The potentials
are plotted for fixed time corresponding to the maximal field amplitude. The
saddles move along the dashed lines when the electric field points in the
positive $z$-direction and along the second obtained by reflections on $z=0$
during the other half of the field cycle. Note that the angle between the
$z$-axis and the dashed line increases with the number of electrons.
}\label{one}
\end{figure}

\begin{figure}[hbt]
\epsfig{file=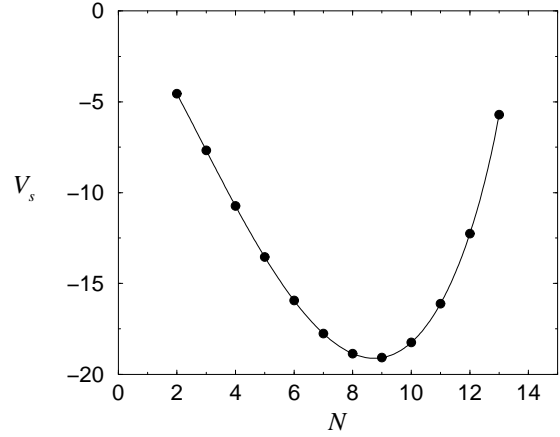%
,scale=0.34,angle=-90
}
\caption[]{
The energy of the saddle point of the potential 
(\protect{\ref{p}}), in the scaled variables (\protect{\ref{sc}}), versus 
number of electrons $N$. The values correspond to the maximal field amplitude. 
}\label{two}
\end{figure}

Equipotential curves of Eq.~(\ref{p}) in the scaled variables (\ref{sc}),
for $N=3$ and $N=6$ are shown in Fig.~\ref{one}. The saddles are located along
the lines $z=r_s\cos\theta_s$ and $\rho=r_s\sin\theta_s$ with $\theta_s=\theta$
or $\theta_s=\pi-\theta$ where
\beq
\theta=\arctan\left[\left(\frac{4N\sin(\pi/N)}{N-1}\right)^{2/3}
-1\right]^{-1/2}
\label{th}
\eeq
and
\beq
r_s^2=\frac{N}{f_t}\sqrt{1-\left(\frac{N-1}{4N\sin(\pi/N)}\right)^{2/3}}
\label{rs}
\eeq
with $f_t=|f(t)\cos (\omega t+\phi)|$. The energy of the saddle is 
\beq
V_s=-\sqrt{4N^3f_t}
\left[1-\left(\frac{N-1}{4N\sin(\pi/N)}\right)^{2/3}\right]^{3/4}.
\label{se}
\eeq
During a field cycle the saddle moves in from infinity along the line
$\theta_s=\theta$, back out to infinity and then in and out again along
the line $\theta_s=\pi-\theta$. The angle $\theta$ increases with the number of
electrons $N$, while the energy of the saddle changes non-monotonically 
as shown in
Fig.~\ref{two}. For $N\ge 14$ the saddle disappears. For this
many electrons the repulsion between the electrons is stronger 
than the attraction to the nucleus, see Eq.~(\ref{p}) for $z=0$. The
situation with $N\ge 14$ is rather an extreme case from the experimental 
point of view but it reflects the fact that a bound state for a 
neutral atom with $N$ electrons symmetrically distributed in a 
plane can not be formed if $N\ge 14$.

The saddle which is opened by the external field allows electrons to escape even
if the total energy $E$ of the highly excited complex is negative. It is enough,
classically, that $E$ is greater than the energy of the saddle. The saddle 
thus provides a kind of transition state \cite{Wigner,Pollak}
for the ionization process: once the electrons cross it, they are accelerated 
by the field and pulled further away, making a return rather unlikely. 
Moreover, they can acquire the missing energy so that the electrons can 
escape even when the field vanishes. The field thus plays a double role 
in determining a threshold for this process: during the first stages 
of the rescattering process it provides the energy for the collision 
complex and during the final stages it opens the path for multiple escape.


In the experiment \cite{rottke} of triple (also single and double) ionization 
of Ne by ultrashort (30~fs) laser
pulses at 795~nm and at intensities 1.5~PW/cm$^2$ the distributions 
of momenta of ions parallel and perpendicular to the polarization axis
have been measured.
In the limit of negligibly small momentum transfer by the absorbed photons, 
the ion momentum $\vec p_{ion}$ reflects the sum of the momenta of 
the emitted electrons,
$\vec p_{ion}=-\sum_i \vec p_i$ \cite{weber1,rottke}. 
To calculate the experimental distributions
corresponding to the triple ionization of Ne we have performed 
classical trajectory simulations for $N=3$. Note that the correct
equations of motion for a single electron follow from the 
Hamiltonian (\ref{h}) after division by $N$.

\begin{figure}[hbt]
\psfig{file=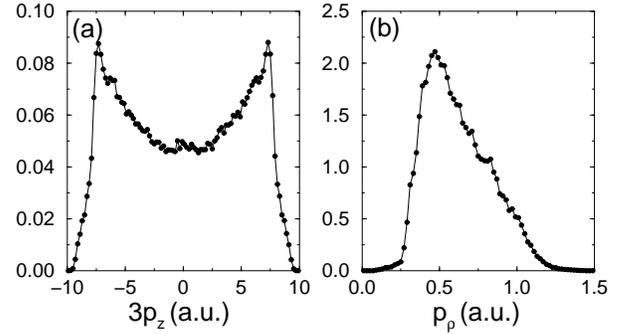%
,scale=0.34,angle=-90
}
\caption[]{
Final distribution of electrons momenta parallel [panel (a)] and perpendicular
[panel (b)] to the field polarization axis for $E=-1$~a.u. 
Panel (a) corresponds to the
experimental distribution of ions momenta measured in the triple ionization of
Ne \protect{\cite{rottke}}. Note that, in panel (b), the distribution rapidly
falls to zero around $p_{\rho}\approx 0$ reflecting the effect of repulsion
between the electrons. The distributions are based
on an ensemble of $2\cdot10^5$ trajectories.
}\label{three}
\end{figure}

The Hamiltonian (\ref{h}) with $N=3$ is the full Hamiltonian for 
a Li with its three electrons. In order to relate it to
triple ionization from atoms with more electrons the zero point
in energy has to be shifted to the threshold for the three
electron continuum, and all energies have to be taken relative
to this level. Moreover, interactions with the electrons in the 
core are neglected. 
Specifically, for the modeling of the experiments 
\cite{rottke} on triple ionization in Ne
we thus assume that in a rescattering process the energy transfer
is less than the threshold for triple ionization (about $4.6$~a.u.).
The precise value of the initial energy $E$ depends on the details
of the rescattering process and can thus not be determined within
our model: it is a free parameter.
The rescattering event is most likely to take place when 
the field amplitude is high, so we have started the simulation 
close to the top 
of the pulse, i.e. for $t_0=0.33T_d$, where $T_d=2412$~a.u. corresponds to 
the experimental pulse duration \cite{rottke}. 
The corresponding amplitude and frequency 
of the field are $F=0.207$~a.u. and $\omega=0.057$~a.u. For such field
parameters the saddle energy equals $-3.5$~a.u. 
We have performed numerical 
simulations for several initial energies. The distributions of the electron
momenta for $E=-1$~a.u. are shown in Fig.~\ref{three}. 
The distribution for
parallel momenta can be compared with the corresponding experimental
distribution of ions momenta \cite{rottke}. 
The main features of the distribution, its width and characteristic 
double hump structure, are well reproduced in our calculation.
The minimum in the distribution is, however, less pronounced than 
that observed in the experiment.
Overall, we take this good
agreement as strong indication that triple ionization can only occur in the
neighborhood of the symmetric process discussed here. 

For initial energies below about $-3$~a.u. the double hump structure 
disappears and only a single maximum remains. 
This corresponds to experiments with weaker fields 
which transfers less energy to the rescattered electron. Then the electrons 
crossing the saddle have smaller kinetic energy and the interaction with the 
remainder of the pulse blurs the distribution. Such a change in the character
of the distribution with decrease of field intensity has also 
been observed in the
experiment of double ionization of He atoms \cite{weber1}. 

The distribution for perpendicular ion momenta can not 
be calculated in our model since this component vanishes 
exactly by symmetry. But we can calculate the 
transverse momentum of an individual escaping electron, as shown in 
Fig.~\ref{three}. 
The absence of events around zero reflects simply the effect of the repulsion
between electrons. Measurements of this quantity should provide further
test of this model.

In conclusion we have shown that the double humped structure in ion momenta
distribution observed in the experiment of triple ionization of Ne 
\cite{rottke} can be calculated within a classical simulation 
assuming symmetrically escaping electrons. This multi-ionization path
can be extended to include up to 13 electrons. The successful comparison with 
the experimental data supports the idea that the dominant contribution
to ionization comes from initial conditions in the highly excited
collision complex that are asymptotic to this 
symmetry plane, very much as in Wanniers analysis \cite{Wannier,Rau,Rost}. 

We would like to thank Harald Giessen for stimulating our interest
in this problem and for discussions of the experiments.
Financial support by the Alexander von Humboldt
Foundation and by KBN under project 2P302B00915
are gratefully acknowledged.


\begin{figure}[hbt]
\epsfig{file=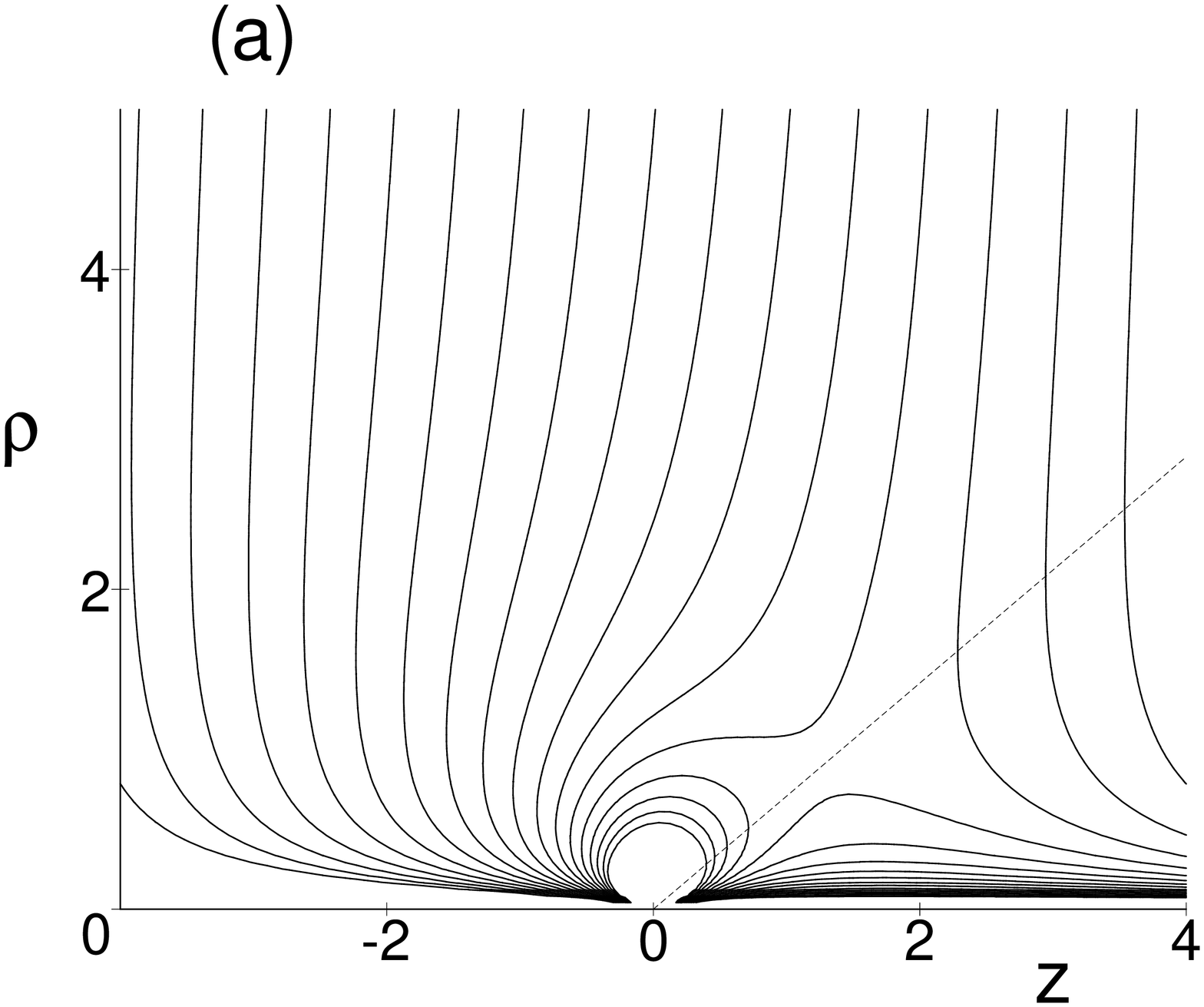%
,scale=0.34,angle=-90
}
\end{figure}

\vfill
\end{multicols}
\end{document}